\documentclass[latex,pdftex]{Interspeech}
\usepackage{bbding}
\usepackage{multicol}
\usepackage{multirow}
\usepackage{graphicx}
\usepackage{url}
\usepackage{CJKutf8} 
\usepackage{comment}
\newcommand\blfootnote[1]{%
  \begingroup
  \renewcommand\thefootnote{}\footnote{#1}%
  \addtocounter{footnote}{-1}%
  \endgroup
}

\interspeechcameraready 

\title{Grapheme-Coherent Phonemic and Prosodic Annotation of Speech \\ by Implicit and Explicit Grapheme Conditioning}

\author[affiliation={1,*}]{Hien}{Ohnaka}
\author[affiliation={2}]{Yuma}{Shirahata}
\author[affiliation={2}]{Byeongseon}{Park}
\author[affiliation={2}]{Ryuichi}{Yamamoto}

\affiliation{}{Nara Institute of Science and Technology}{Japan}
\affiliation{}{LY Corporation}{Japan}
\email{{p-hionaka,yuma.shirahata}@lycorp.co.jp}
\keywords{prosodic annotation, accent estimation, data augmentation, text-to-speech}

\newcommand{\red}[1]{\textcolor{black}{#1}}
\newcommand{\pedit}[1]{\textcolor{black}{#1}}
\newcommand{\sedit}[1]{\textcolor{black}{#1}}
\newcommand{\jp}[1]{ \begin{CJK}{UTF8}{ipxm}#1\end{CJK} }
\usepackage{comment}

\begin{document}

\maketitle

\fontsize{9.0}{10.5}\selectfont

\begin{abstract}
\sedit{We propose a model to obtain phonemic and prosodic labels of speech that are coherent with graphemes.
Unlike previous methods that simply fine-tune a pre-trained ASR model with the labels, the proposed model conditions the label generation on corresponding graphemes by two methods:
1) Add implicit grapheme conditioning through prompt encoder using pre-trained BERT features.
2) Explicitly prune the label hypotheses inconsistent with the grapheme during inference.
These methods enable obtaining parallel data of speech, the labels, and graphemes, which is applicable to various downstream tasks such as text-to-speech and accent estimation from text.
Experiments showed that the proposed method significantly improved the consistency between graphemes and the predicted labels. Further, experiments on accent estimation task confirmed that the created parallel data by the proposed method effectively improve the estimation accuracy.}
\end{abstract}

\section{Introduction}
\blfootnote{\red{* Work done during an internship at LY Corporation.}}
The field of text-to-speech (TTS) \pedit{has advanced significantly} through data-driven approaches based on deep neural networks (DNNs)~\cite{Tacotron,FastSpeech2,VITS}. For training high-quality and diverse-styled TTS models, a large amount of text-speech paired data is required~\cite{naturalspeech2,MegaTTS}. However, since \pedit{manually preparing} text transcriptions for a \pedit{large} amount of unlabeled speech \pedit{samples is costly, some research has generated} transcriptions using automatic speech recognition (ASR) models~\cite{le2024voicebox,speech_machine_chain,speech_machine_chain2}. This approach is particularly effective to languages such as English, where grapheme sequences from ASR represent the reading with high accuracy.

On the other hand, there are some languages\pedit{, like Japanese and Chinese, where one grapheme sequence has multiple readings and accents}. \sedit{For such languages, phonemic and prosodic labels (hereinafter, TTS labels) are typically used as the textual feature~\cite{prosodic_features_TTS_JP1,prosodic_features_TTS_JP2,prosodic_features_TTS_Mandarin}, leading to two challanges. First, it is costly to prepare a sufficient number of TTS labels to train a speech synthesis model. Second, since graphemes are used as input during inference, another model is required that predicts TTS labels from graphemes. Specifically, textual accent estimation, which is a task of prosodic label estimation from graphemes, is important and various models have been proposed~\cite{Jack_accent_estimation,Hida2022accent,kurihara2024accent}. For training these models, paired data of TTS labels and graphemes is required.}

To address \sedit{the first challenge, a method that effectively obtains the TTS labels of speech samples} has been proposed~\cite{pa-whisper}. The key idea of the method is fine-tuning the Whisper ASR~\cite{whisper} using a limited amount of \pedit{the} TTS label dataset. This fine-tuned annotation model can directly acquire TTS labels from speech samples.
Applying this annotation model to a large-scale corpus~\cite{ReazonSpeech,YODAS} can be expected to significantly scale up the data. 
\sedit{On the other hand, the method was suboptimal for the second challenge, i.e., preparing the paired data of TTS labels and graphemes. This is because the method does not incorporate the corresponding grapheme into the prediction of the TTS labels, resulting in labels that are inconsistent with graphemes. } 

\sedit{To address this issue}, we propose a TTS label annotation model conditioned on graphemes. Our approach involves fine-tuning an encoder-only speech foundation model~\cite{OWSM-CTC} with a newly added \textit{implicit} grapheme-conditioning module. Specifically, the module effectively utilizes the prompt encoder in \cite{OWSM-CTC} and \pedit{the} pre-trained BERT features~\cite{BERT,DistilBERT}.
Furthermore, to \textit{explicitly} condition the TTS label prediction on graphemes, we propose a decoding strategy that prunes TTS label hypotheses that are inconsistent with the grapheme, using an external grapheme-to-phoneme (g2p) dictionary. Thanks to these implicit and explicit grapheme conditioning methods, the model is expected to be capable of predicting TTS labels that align with the given graphemes and speech samples. 
In terms of applications, since the proposed method can prepare the parallel data of speech, grapheme, and TTS labels, it can be applied to not only TTS, but also textual accent estimation, among others~\cite{PnG_BERT,Phoneme_BERT_with_Grapheme_pred,Miipher}.

In our experiments, we demonstrate that the proposed annotation model achieves \pedit{a} higher grapheme-to-phoneme match rate compared to baseline models without grapheme conditioning, while maintaining \sedit{the accuracy of phonemic and prosodic labels.}
We also show that utilizing the proposed method for data augmentation in the textual accent estimation task~\cite{Jack_accent_estimation} leads to improved estimation accuracy.

\section{Problem fomulation}
The goal of the annotation is to obtain the parallel data of a TTS label sequence $\bm{y}=\{y_m \in \mathcal{Y}\}_{m=1}^M$ and a grapheme sequence $\bm{g}=\{g_l \in \mathcal{G}\}_{l=1}^L$, and a speech sample $\bm{X}=\{x_n \in \mathbb{R}^{D_\mathrm{in}}\}_{n=1}^N$ from $\bm{X}$. 
Here, $\mathcal{Y}$ and $\mathcal{G}$ represent the vocabularies of TTS label and grapheme tokens, respectively.
$M$,$L$, and $N$ denote the lengths of each sequence. 
$D_\mathrm{in}$ denotes the dimension of the input speech features.
\\\textbf{NLP-based approach:} 
\sedit{A simple approach to obtain a paired data of $\bm{g}$ and $\bm{y}$ is to apply an ASR model to obtain $\bm{g}$ from $\bm{X}$, and then use a natural language processing (NLP)-based methods to obtain $\bm{y}$ from $\bm{g}$~\cite{Jack_accent_estimation,oura2010japanese}}.
Here, the following conditional dependency assumptions are introduced:
\begin{align}
    p(\bm{y}, \bm{g}|\bm{X})&=p(\bm{y}_\mathrm{ph},\bm{y}_\mathrm{ps}, \bm{g}|\bm{X}) \nonumber\\
    &=p(\bm{y}_\mathrm{ph},\bm{y}_\mathrm{ps}|\bm{g})p(\bm{g}|\bm{X}), \label{eq:nlp-based}
\end{align}
where $\bm{y}_\mathrm{ph}$ and $\bm{y}_\mathrm{ps}$ denote phonemic label sequence and prosodic label sequence, respectively.
In Eq. (\ref{eq:nlp-based}), the first term \sedit{represents the NLP-based method, which is independent of the speech $\bm{X}$.}
The second term corresponds to the ASR, which can be easily optimized by leveraging existing high-quality ASR models~\cite{whisper,nue-asr}.
\sedit{The problem of this approach is that the speech and TTS labels may mismatch since speech $\bm{X}$ is ignored in the TTS label prediction. This is caused by the one-to-many relationship between graphemes and TTS labels} (e.g., the word ``\jp{化学}'' can be read as ``ka ga ku'' or ``ba ke ga ku'').
\begin{figure}[t!]
\begin{center}
    \includegraphics[width=0.85\linewidth]{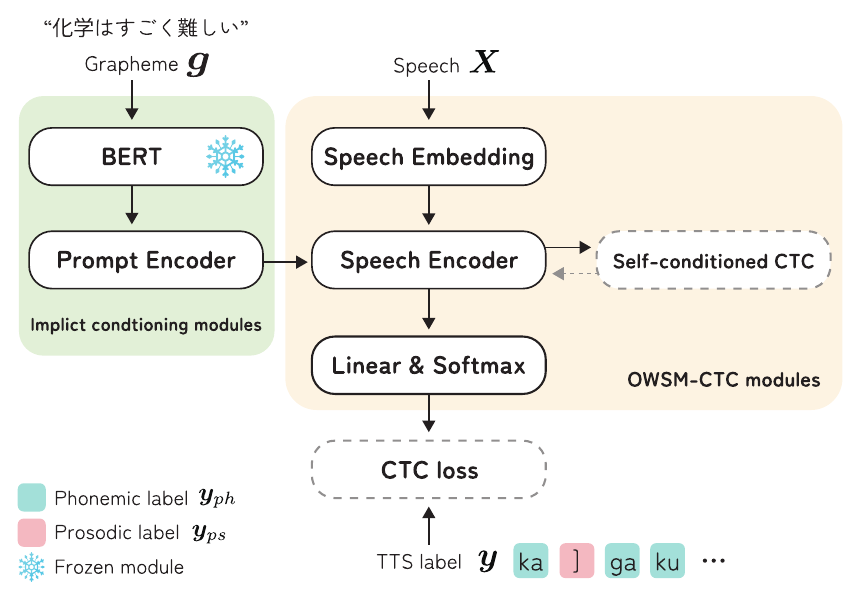}
    \vspace{-2.5mm}
  \caption{Architecture of the proposed model.}
  \label{fig:model_architecture}
\end{center}
\vspace{-9.5mm}
\end{figure}
\\\textbf{Previous annotation model~\cite{pa-whisper}:}
To address this one-to-many mapping problem, 
\cite{pa-whisper} trained an annotation model to directly predict TTS labels from speech samples using a paired dataset of $(\bm{X},\bm{y})$.
During inference, the model predicts the TTS label sequence $\hat{\bm{y}}$ based on the given speech sample $\bm{X}$ as follows:
\begin{align}
    \hat{\bm{y}} = \underset{y \in \mathcal{Y}^*}{\mathrm{argmax}}~ p(\bm{y}|\bm{X}). \label{eq:infer_pa-whisper}
\end{align}
Here, $\mathcal{Y}^*$ is corresponding a set of all possible hypotheses.
\sedit{Since reading and prosody are expressed in speech,} the direct prediction of $\hat{\bm{y}}$ from $\bm{X}$ solves the one-to-many mapping problem. \sedit{For the prediction of $\bm{g}$, this method need to use an ASR model independently from the TTS label prediction:}
\begin{align}
    \hat{\bm{g}} = \underset{g \in \mathcal{G}^*}{\mathrm{argmax}}~ p(\bm{g}|\bm{X}). \label{eq:infer_asr}
\end{align}
Here, $\mathcal{G}^*$ denotes a set of all possible hypotheses. \sedit{Due to the independent prediction of $\bm{y}$ and $\bm{g}$, the generated TTS labels and graphemes tend to have many mismatched samples. As a result, it is difficult to apply them to tasks such as textual accent estimation.}
\\\textbf{Proposed method:}
To address the mismatch between TTS labels and graphemes,
we propose an annotation model that is conditioned not only on speech but also on graphemes. 
The generative model in the proposed method is expressed by the following equation:
\begin{align}
p(\bm{y},\bm{g}|\bm{X})&=p(\bm{y}|\bm{X},\bm{g})p(\bm{g}|\bm{X}). \label{eq:modeling_proposed_method}
\end{align}
Here, the first term corresponds to the proposed grapheme-conditioned annotation model, and the second term corresponds to the ASR model.
Training requires paired data $(\bm{X},\bm{y},\bm{g})$. Although $\bm{g}$ is additionally required for training, given the availability of high-quality ASR models, the difficulty of data construction is as low as the aforementioned method~\cite{pa-whisper}.
During inference, $\hat{\bm{y}}$ is obtained using the following equation:
\begin{align}
    \hat{\bm{y}} = \underset{y \in \mathcal{Y}^*}{\mathrm{argmax}}~ p(\bm{y}|\bm{X},\bm{g})p(\bm{g}|\bm{X}). \label{eq:infer_prop}
\end{align}
When using the predicted grapheme $\hat{\bm{g}}$ from ASR in the second term, TTS labels that align with both the speech and the predicted grapheme $\hat{\bm{g}}$ are obtained.
Additionally, when access to the ground-truth grapheme $\bm{g}$ is possible, such as in speech recording, the prediction of the TTS label can also consider that ground-truth grapheme.

\section{Proposed method}

\subsection{Model architecture} \vspace{-1mm}
In the proposed method, we adopted OWSM-CTC~\cite{OWSM-CTC} as the base pre-trained model. 
OWSM-CTC is an encoder-only speech foundation model pre-trained on large-scale multilingual data, including Japanese, for ASR and speech translation tasks. 
As shown in Fig.~\ref{fig:model_architecture}, the model consists of a \sedit{speech embedding layer}, a speech encoder, and a prompt encoder. 
For the Speech encoder, CTC loss~\cite{CTC_loss} calculation and conditioning were applied at intermediate layers based on Self-conditioned CTC~\cite{self_conditioned_CTC}. 
The Prompt encoder is pre-trained for long-form ASR, where graphemes serving as prefixes of the input speech are provided as conditioning.
Unlike \cite{pa-whisper}, which adopted Whisper-small as the base model, we adopted OWSM-CTC for the following reasons: 1) robustness to repetition errors in the non-autoregressive model, 2) good generalization performance with a large number of parameters, and 3) the effective utilization of the Prompt encoder for implementing implicit conditioning, as discussed in Sec.~\ref{sec:impli_cond}.

\subsection{Implicit conditioning} \label{sec:impli_cond}
\vspace{-1mm}
In realizing the annotation model based on speech and graphemes as described in Eq.~(\ref{eq:modeling_proposed_method}), the vast variety of graphemes is a challenge. 
Graphemes exhibit a wide range of variations compared to phonemes, making it difficult to prepare paired data with speech that covers all these variations.

To address this issue, we introduce implicit conditioning by leveraging a pre-trained BERT~\cite{BERT,DistilBERT} and a Prompt encoder~\cite{OWSM-CTC}. 
Conditioning is achieved by serially connecting the BERT module and the Prompt encoder as shown in Fig.~\ref{fig:model_architecture}.
The knowledge obtained from BERT's pre-training on a large text corpus is expected to enable adaptation to a wide range of domains, even with limited data~\cite{BERT_CTC}.
Additionally, since the Prompt encoder is pre-trained for prefix grapheme embeddings in long-form ASR, it is considered useful as a base module for grapheme conditioning.

\subsection{Explicit conditioning}
\vspace{-1mm}
\label{sec:expli_cond}
\begin{figure}[t!]
\vspace{-6mm}
\begin{center}
    \includegraphics[width=0.87\linewidth]{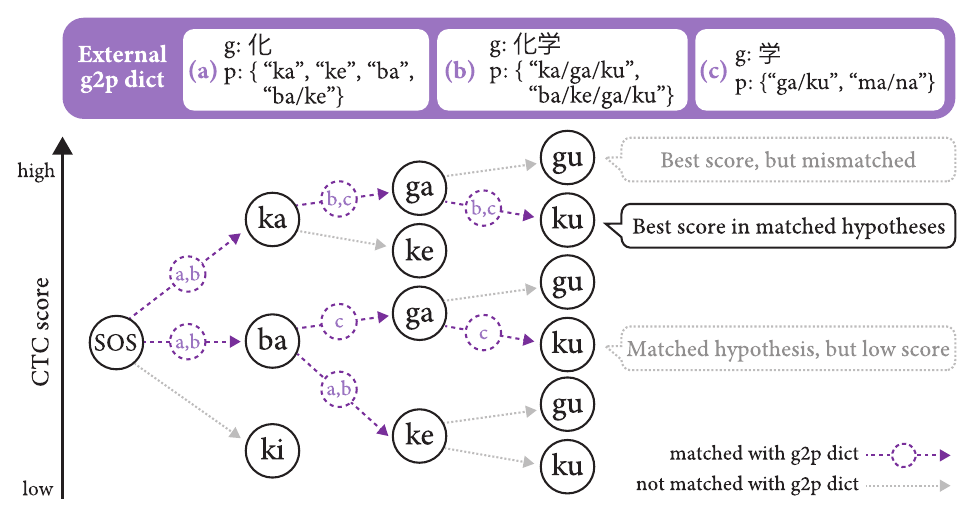}
    \vspace{-2mm}
  \caption{
Concept of explicit conditioning. 
\sedit{The matching between phonemes and graphemes is checked phoneme-by-phoneme using an external g2p dictionary, and the best-scored hypothesis among the matched ones is selected.}
The figure illustrates an example where the graphemes correspond to ``\jp{化学}'' (Chemistry).
}
\vspace{-6mm}
  \label{fig:decoding_concept}
\end{center}
\vspace{-3mm}
\end{figure}
Preliminary experiments have confirmed that applying implicit conditioning alone cannot fully address minor errors, such as consonant confusions~\cite{consonant_confusions}.
To correct these minor errors post hoc and further enhance the matching between graphemes and phonemes, we introduce a decoding strategy with an external g2p dictionary.
The proposed decoding concept named explicit conditioning is presented in Fig.~\ref{fig:decoding_concept}.
Firstly, to reduce the speed down of inference, we extract only the probability features in time indices where ``\verb|<|blank\verb|>|'' is not the top token.
Secondly, we perform \pedit{a} hypothesis search from the beginning, following a procedure in greedy search.
At each step, partial matches between graphemes and phonemic sequences within the TTS label are checked using an external g2p dictionary and dynamic programming~\cite{DP}. Finally, the token with the highest score among the matching tokens is retained. 
Here, prosodic labels are always considered to be a match.
Note that, within the proposed method, phonemes that are clearly different can be eliminated using the CTC score, so it is only sufficient for the dictionary to have high recall (low precision does not matter).
Therefore, the difficulty of constructing the dictionary itself is low enough.
\red{The phonemes in the final TTS label sequence are expected to align with the entire grapheme sequence.}
At this point, the TTS label prediction in Eq.~(\ref{eq:infer_prop}) is replaced as follows:
\begin{align}
    \hat{\bm{y}} = \underset{y \in \tilde{\mathcal{Y}}^*}{\mathrm{argmax}}~p(\bm{y}|\bm{X},\bm{g})p(\bm{g}|\bm{X}).
\end{align}
Here, $\tilde{\mathcal{Y}}^*$ is the set of hypotheses whose matches are ensured by the external dictionary.
 
\vspace{-1mm}
\section{Experiments}
\subsection{Evaluation of proposed annotation model}
\vspace{-1mm}
\subsubsection{Experimental setup}
\vspace{-1mm}
\textbf{Datasets.}
In training the proposed model, we adopted the following two dataset conditions, similar to previous work~\cite{pa-whisper}:
1) For model construction from a limited amount of labeled data, we used the publicly available Japanese speech corpus JSUT~\cite{JSUT}, which consists of utterances from a single female speaker. 
Specifically, we utilized the basic5000 subset along with its manually annotated TTS labels~\footnote{\url{https://github.com/sarulab-speech/jsut-label}}. 
This subset comprises $5,000$ text samples and $6.78$ hours of speech, which was divided into $4,413$ samples for training and $248$ samples for validation.
2) To evaluate the performance of the proposed model itself, we constructed a model from a larger dataset using our proprietary Japanese speech corpus, which includes recordings from six male and eleven female speakers with manual labels. 
This corpus consists of $173,987$ samples and $207.96$ hours of speech, divided into $153,551$ samples for training, $4,449$ for validation, and $14,000$ for testing. 
We call this dataset LARGE in the following sections.
\\\textbf{TTS data augmentation.}
Similar to previous research~\cite{pa-whisper}, we applied TTS data augmentation to address the limited data availability in the JSUT dataset.
This data augmentation was conducted through the following steps:
1) We applied NLP-based phonemic/prosodic label prediction to a text corpus to obtain pseudo TTS labels $\hat{\bm{y}}$.
We used phonemic label estimation with Mecab~\cite{mecab} and the Japanese dictionary Unidic~\cite{unidic}, along with textual accent estimation using an accent estimator~\cite{Jack_accent_estimation} trained on the clean data for fine-tuning described in Sec.~\ref{sec:setup_accent_estimation}.
2) We obtained synthetic speech $\hat{\bm{X}}$ from the pseudo TTS labels to create paired data $(\hat{\bm{X}},\hat{\bm{y}})$.
We used Period VITS~\cite{period_vits}, configured as in previous study~\cite{pa-whisper}.
Data augmentation was applied to the LARGE text dataset.
Additionally, to further expand the grapheme domain, we performed similar data augmentation using transcribed text from ReazonSpeech~\cite{ReazonSpeech} (small) with OWSM-CTC~\cite{OWSM-CTC} v3.1~\footnote{\url{https://huggingface.co/espnet/owsm_ctc_v3.1_1B}}.
In JSUT, both types of data augmentation were applied, while in LARGE, only the latter was used.
Furthermore, to address noisy speech, we applied noise-reverberation data augmentation to each sample. 
This involved adding noise from the DEMAND noise database~\cite{DEMAND} at an SNR of $0$ to $10$ dB and applying reverberation using one of the RIRs from the ACE challenge~\cite{ACE_challenge} with an 80\%.
\\\textbf{TTS label.}
We adopted the definition~\cite{pa-whisper} based on the design by Kurihara et al~\cite{prosodic_features_TTS_JP1}.
The prosodic status of each mora is represented by five labels, considering the Japanese pitch accent rules of the Tokyo dialect: (1) Pause ``\_'', (2) Accent change from low to high ``['', (3) Accent change from high to low ``]'', (4) Accent phrase boundary ``\#'', and (5) Raise-type boundary pitch movement ``?''.
Phonemic labels were represented using Katakana characters corresponding to each mora. 
Following the success of previous studies~\cite{pa-whisper,mixed_seq1,mixed_seq2,mixed_seq3}, we adopted a mixed sequence of phonemic and prosodic labels as shown in Fig.~\ref{fig:model_architecture}.
\\\textbf{Model configuration.}
We adopted OWSM-CTC~\cite{OWSM-CTC} v3.1 as the base model. 
During training, the weights of the first 5 layers of the 27-layer Speech Encoder were fixed, while all other parameters were updated. 
Training was conducted with a batch size of $32$ for $50,000$ steps, with validation performed every $1,000$ steps to determine the best weights based on the lowest sequence error rate. 
For grapheme embeddings, we used line-distil-bert-base-japanese\footnote{\url{https://huggingface.co/line-corporation/line-distilbert-base-japanese}}~\cite{DistilBERT}. 
\red{
BERT and prompt encoder were connected with a linear layer to align the dimensions. 
}
Other conditions were consistent with those used in pre-training.
\\\textbf{External grapheme-to-phoneme dictionary.}
Based on mpaligner~\cite{mpaligner_apsipa,mpaligner_interspeech}, we obtained many-to-many alignment results from grapheme-phoneme paired data.
For the paired data, we used $2,188,937$ samples, including proprietary data and naist-jdic~\footnote{\url{https://ja.osdn.net/projects/naist-jdic/}}.
Then, each sample was parsed into minimum unit to create the g2p dictionary. An example of the entry is like $\{$``\jp{化学}'': ``ka ga ku'', ``ba ke ga ku''$\}$.
The total number of keys in the dictionary was $17,278$, with an average of $2.55$ partial phonemes per key.

\vspace{-2mm}
\subsubsection{Experimental evaluation and results}
\vspace{-1mm}
We evaluated the performance of the proposed annotation method using the Phoneme Error Rate (PER) and the F$_1$ score of prosodic labels \sedit{(Prosody $F_1$)}~\cite{pa-whisper} on $14,000$ samples from the LARGE test set.
Additionally, to assess the consistency of phoneme and grapheme, the match rate between graphemes and predicted TTS labels (G2P match) was calculated.
Five phoneme estimation candidates were generated from graphemes using Mecab~\cite{mecab}, and a match was determined if the phonemic sequence of the TTS label matched any of these candidates.
For this evaluation, $5,322$ samples from ReazonSpeech~\cite{ReazonSpeech} (tiny) were also used.
ReazonSpeech is a noisy dataset that includes factors such as noise and intrusion from external speakers.
\begin{table}[t!] 
\caption{Overall experimental results are presented, showing the grapheme-to-phoneme match rate (G2P match) for transcribed graphemes using OWSM-CTC and the predicted TTS labels from each method, Phoneme Error Rate (PER), and Prosody F$_1$ score (Pros. F$_1$) under two dataset conditions.} \label{tab:overall_result}
\vspace{-3mm}
{\fontsize{6pt}{0.3cm}\selectfont
\begin{tabular}{cc|ccc} 
\hline \hline
Dataset & 
\begin{tabular}[c]{@{}c@{}}
Graph.\\cond.\\
(Prop.)
\end{tabular} & 
\begin{tabular}[c]{@{}c@{}}
G2P match ($\uparrow$)\\
(LARGE/Reazon)
\end{tabular}& PER ($\downarrow$) & Pros. F$_1$ ($\uparrow$) \\ \hline 
JSUT
& -
& $53.70$/$16.99$\%
& $0.28$\%              
& $91.06$\%   \\                                                               
JSUT-TTSaug
& -                                         
& $38.55$/$8.01$\%
& $\mathbf{0.21}$\%              
& $94.63$\%    \\                                                              
JSUT-TTSaug
& \Checkmark              
& $\mathbf{81.47}$/$\mathbf{52.25}$\% 
& $1.12$\%              
& $\mathbf{95.12}$\%     \\       \hline     
LARGE                         
& -                                         
& $71.04$/$15.33$\%
& $\mathbf{0.08}$\%              
& $\mathbf{98.89}$\%          \\                                                        
LARGE-TTSaug
&  -                                         
& $71.38$/$21.06$\%                                                        
& $0.15$\%              
& $98.30$\% \\                                                                 
LARGE-TTSaug
& \Checkmark                  
& $\mathbf{82.20}$/$\mathbf{57.27}$\%                                                                             
& $0.93$\%              
& $98.08$\%  \\ \hline \hline 
\end{tabular}
}\vspace{-5.5mm}
\end{table}
\\\textbf{Overall performance assessment.}
Table~\ref{tab:overall_result} presents the evaluation results under the two dataset conditions.
In this evaluation, transcribed text from OWSM-CTC v3.1 was used for grapheme conditioning.
Similar to results from previous research~\cite{pa-whisper}, introducing TTS data augmentation in the limited JSUT dataset improved both the PER and Prosody F$_1$ score.
Furthermore, in the LARGE dataset, using data augmentation on ReazonSpeech (tiny) led to an improvement in the G2P match rate, albeit with a slight degradation in PER and Prosody F$_1$ score on the original domain's LARGE test set. 

It was observed that utilizing grapheme conditioning significantly improved the match rate in both conditions, confirming the effectiveness of the proposed method.
The Prosody F$_1$ score was comparable to or better than the baseline and the case without grapheme conditioning, indicating that the match rate was improved without affecting the accuracy of prosodic label prediction. 
On the other hand, the PER worsened with grapheme conditioning.
This degradation is due to transcription errors present in the graphemes themselves, which induce additional errors.
This issue can be resolved with improvements in the ASR model itself, suggesting that it is not a major concern within the scope of this study.
\begin{table}[t!] 
\caption{The result utilizing ground-truth graphemes.} \label{tab:ablation_gt_grapheme}
\vspace{-3mm}
{\fontsize{7pt}{0.28cm}\selectfont
\begin{tabular}{c|ccc}
\hline \hline
\begin{tabular}[c]{@{}c@{}}
Grapheme\\type
\end{tabular}
 & 
\begin{tabular}[c]{@{}c@{}}
G2P match ($\uparrow$)\\
(LARGE)
\end{tabular}& PER ($\downarrow$) & Pros. F$_1$ ($\uparrow$) \\ \hline 
\multicolumn{4}{c}{JSUT-TTSaug w/ Grapheme conditioning} \\
Ground-truth grapheme                                    
& $\mathbf{90.13}$\% 
& $\mathbf{0.26}$\%              
& $95.00$\%      \\                                                            
OWSM-CTC~\cite{OWSM-CTC}
& $81.47$\% 
& $1.12$\%              
& $\mathbf{95.12}$\%  \\ \hline     
\multicolumn{4}{c}{LARGE-TTSaug w/ Grapheme conditioning} \\
Ground-truth grapheme
& $\mathbf{91.17}$\% 
& $\mathbf{0.09}$\%              
& $\mathbf{98.62}$\%  \\                                                                
OWSM-CTC~\cite{OWSM-CTC}                                        
& $82.20$\%     
& $0.93$\%              
& $98.08$\% \\ \hline \hline  
\end{tabular}
}\vspace{-2mm}
\end{table}
\\\textbf{Performance with ground-truth graphemes.}
To investigate the effect of ASR transcription errors, we conducted experiments using ground-truth grapheme from the LARGE test set for conditioning. Table~\ref{tab:ablation_gt_grapheme} presents the results.
In this case, it is evident that phonemic labels can be predicted with accuracy comparable to or better than the condition without grapheme conditioning in Table~\ref{tab:overall_result}. 
In situations where only speech and graphemes are available in a speech corpus, and the goal is to annotate TTS labels, access to ground-truth graphemes is possible. 
These results demonstrate the utility of the proposed method under such conditions.
\begin{table}[t!]
\caption{Ablation study results on the two conditioning methods.} \label{tab:ablation_impact_prop}
\vspace{-3mm}
{\fontsize{7pt}{0.28cm}\selectfont
\begin{tabular}{cc|ccc}
\hline \hline
\begin{tabular}[c]{@{}c@{}}
Implicit\\cond.
\end{tabular}
 & 
\begin{tabular}[c]{@{}c@{}}
Explicit\\cond.
\end{tabular} & 
\begin{tabular}[c]{@{}c@{}}
G2P match ($\uparrow$)\\
(LARGE/Reazon)
\end{tabular}& PER ($\downarrow$) & Pros. F$_1$ ($\uparrow$) \\ \hline 
\multicolumn{5}{c}{JSUT-TTSaug} \\
-
& -
& $38.55$/$8.01$\%
& $\mathbf{0.21}$\%              
& $94.63$\% \\                                                                 
-
& \Checkmark                                      
& $71.80$/$25.51$\%
& $0.61$\%              
& $93.95$\% \\                                                                 
\Checkmark                         
& -                                       
& $73.38$/$36.31$\% 
& $0.58$\%              
& $\mathbf{95.41}$\%  \\                                                                
\Checkmark
& \Checkmark
& $\mathbf{81.47}$/$\mathbf{52.25}$\% 
& $1.12$\%              
& $95.12$\%  \\ \hline                                                           
\multicolumn{5}{c}{LARGE-TTSaug} \\
-
& -
& $71.38$/$21.06$\%                                                        
& $\mathbf{0.15}$\%              
& $98.30$\%     \\                                                             
-
& \Checkmark                                      
& $77.76$/$36.69$\%
& $0.61$\%              
& $\mathbf{98.38}$\%     \\                                                             
\Checkmark
& -
& $80.28$/$46.57$\% 
& $0.55$\%              
& $98.18$\%   \\                                                               
\Checkmark                         
& \Checkmark                                         
& $\mathbf{82.20}$/$\mathbf{57.27}$\%     
& $0.93$\%              
& $98.08$\%   \\ \hline \hline    
\end{tabular}
}\vspace{-5mm}
\end{table}
\\\textbf{Effectiveness of the grapheme conditioning methods.}
We conducted an ablation study of two grapheme conditioning methods described in Sec.~\ref{sec:impli_cond} and \ref{sec:expli_cond}.
The results are shown in Table~\ref{tab:ablation_impact_prop}.
In both dataset conditions, adopting either method alone significantly improved the G2P match rate, and employing both methods further amplified this effect.
These results suggest that the two methods do not compete with each other and that each addresses different mismatches.

\vspace{-1mm}
\subsection{Application to textual accent estimation}
\vspace{-1mm}
To confirm the usefulness of the grapheme-TTS label parallel data obtained by the proposed method, we trained a textual accent estimation model, which requires the parallel data.\footnote{
\red{Although TTS experiments were not conducted in this paper, the results in Table 3 and \cite{pa-whisper} indicates that the proposed method is also effective for TTS applications.}
}
Specifically, we first pre-trained the model with a large amount of augmented data obtained by the proposed method, and then fine-tuned it with manually-labeled data.

\vspace{-1mm}
\subsubsection{Experimental setup} \label{sec:setup_accent_estimation}
\vspace{-1mm}
\textbf{Dataset.}
For the augmented data in pre-training, we used ReazonSpeech (large)~\cite{ReazonSpeech} as the target speech corpus, from which TTS labels were obtained. 
This corpus consists of $3,097,590$ audio samples totaling $5,000$ hours. 
As the grapheme condition, we employed transcriptions generated by OWSM-CTC v3.1.
In the preprocessing stage of pre-training, data cleaning was performed and only the samples with g2p match were used.

For fine-tuning with clean data, we used a proprietary dataset annotated by linguistic experts, which includes graphemes and corresponding TTS labels.
The dataset comprises $93,386$ samples, divided into train/val/test sets with $84,386$/$4,500$/$4,500$ samples, respectively.
\\\textbf{Model configuration.}
As the framework for the textual accent estimator, we adopted the model by Park et al.~\cite{Jack_accent_estimation}.
This model incrementally predicts \sedit{a group of prosodic labels, i.e.,} intonation phrases (IPs), accent phrases (APs), and accent nucleus (ANs), from grapheme features and mora information (phonemic labels).
Training was conducted for 20 epochs for both pre-training and fine-tuning, and the best weights were selected based on the validation set.
Other conditions were consistent with those in Park et al.~\cite{Jack_accent_estimation}.
\\\textbf{Comparison models.}
We conducted four different approaches to obtain parallel data for pre-training:
\begin{itemize} 
\vspace{-0.5mm}
    \item \textit{w/o pre-training}: A baseline model that skipped the pre-training with parallel data.
    \item \textit{NLPaug}: Another baseline model that prepared parallel data by performing textual accent and reading estimation using the \textit{w/o pre-training} model.
    \item \textit{JSUT-TTSaug}: Prepared parallel data using JSUT-TTSaug w/ Graph. cond. in Table~\ref{tab:overall_result}.
    \item \textit{LARGE-TTSaug}: Prepared parallel data using LARGE-TTSaug w/ Graph. cond. in Table~\ref{tab:overall_result}.
\vspace{-1.5mm}
\end{itemize} 

\vspace{-1mm}
\subsubsection{Impact of data augmentation by proposed method}
\vspace{-1mm}
\begin{table}[t!] 
\caption{The results of application to textual accent estimation.} \label{tab:accent_estimation_result}
\vspace{-3mm}
{\fontsize{6pt}{0.28cm}\selectfont
\begin{tabular}{c|cccc}
\hline \hline
Method &
\begin{tabular}[c]{@{}c@{}}
Number of\\valid samples\\(in pre-training)
\end{tabular} & 
AP ($\uparrow$) &
AN ($\uparrow$) &
AP+AN ($\uparrow$) \\ \hline
\textit{w/o pre-training}
& -
& $88.77$\%
& $72.62$\%
& $71.18$\%\\ 
\textit{NLPaug}
& $2,929,845$ 
& $89.57$\%
& $75.44$\%              
& $74.24$\% \\  
\textit{JSUT-TTSaug}
& $1,471,885$                                   
& $89.45$\%
& $74.03$\%              
& $72.87$\% \\      
\textit{LARGE-TTSaug}                        
& $1,607,287$                                    
& $\mathbf{90.62}$\% 
& $\mathbf{76.49}$\%              
& $\mathbf{75.44}$\%  \\  \hline \hline    
\end{tabular}
}\vspace{-7mm}
\end{table}
To evaluate the effectiveness of data augmentation, we used $2,484$ samples of proprietary test data from a different domain.
The evaluation metrics included sentence-level accuracy for AP, AN, and their combination (AP+AN).

The experimental results are presented in Table~\ref{tab:accent_estimation_result}. 
It can be seen that \textit{LARGE-TTSaug} achieves the best scores, despite having about half the number of samples compared to \textit{NLPaug}. 
This suggests that while \textit{NLPaug} primarily expanded only the grapheme aspect without significantly increasing accent variation, \textit{LARGE-TTSaug} obtained extended data from speech, resulting in a greater effect in terms of accent diversity.
Although improvements were observed with \textit{JSUT-TTSaug}, the improvement was smaller. This is likely influenced by the accuracy of \sedit{the annotation} shown in Table~\ref{tab:overall_result}. 
Since ReazonSpeech, used in this experiment, is relatively noisy, using a cleaner TTS-specific dataset might improve results. Evaluating this remains a task for future research.

\vspace{-1mm}
\section{Conclusion}
\vspace{-1mm}
We proposed the annotation model from speech using grapheme conditioning. 
Applying this method to downstream tasks beyond textual accent estimation is a challenge for future work.

\bibliographystyle{IEEEtran}
\bibliography{mybib}

\end{document}